\documentclass[a4paper,11pt]{article}
\usepackage{graphicx,rotating,slashed,amsmath,xcolor,amssymb,amsfonts,colortbl,cite,subcaption}
\usepackage[totalheight = 23cm, totalwidth = 17cm]{geometry}
\usepackage{physics}
\usepackage{jheppub,amscd,braket,amsthm}
\usepackage{bm}
\numberwithin{equation}{section}

\definecolor{bluc}{cmyk}{1,1,0,0.1}
\definecolor{rossoCP3}{cmyk}{0,.88,.77,.40}
\definecolor{rosso}{cmyk}{0,1,1,0.4}
\definecolor{rossos}{cmyk}{0,1,1,0.55}
\definecolor{rossoc}{cmyk}{0,1,1,0.2}
\definecolor{verdes}{cmyk}{0.92,0,0.59,0.4}

\hypersetup{colorlinks, bookmarksopen, bookmarksnumbered,
citecolor=verdes, linkcolor=bluc, pdfstartview=FitH, urlcolor=rossos}

\usepackage{comment}

\usepackage{mdframed}

\newcommand{\mLb}{\mathcal{L}_{\beta}}

\newcommand{\mD}{\mathcal{D}}

\newcommand{\bea}{\begin{eqnarray}}
\newcommand{\eea}{\end{eqnarray}}
\newcommand{\be}{\begin{equation}}
\newcommand{\ee}{\end{equation}}

\newcommand{\phia}{\phi_{\rm a}}

\newcommand{\ib}{\iota_{\beta}}

\definecolor{purpleheart}{rgb}{0.41, 0.21, 0.61}

\definecolor{limegreen}{rgb}{0.2, 0.8, 0.2}

\newcommand{\ud}{\,\mathrm{d}} 

\title{Emergent structures in open 
EFTs}

\author[a,b]{Perseas Christodoulidis}
\affiliation[a]{Department of Science Education, Ewha Womans University, Seoul 03760, Republic of Korea}
\affiliation[b]{Korea Institute for Advanced Study, Seoul 02455, Republic of Korea}
\emailAdd{perseas@ewha.ac.kr}

\abstract{
Open effective field theories provide a systematic framework for describing systems coupled to an environment, where dissipation, noise, and modified conservation laws naturally arise. Working within the Schwinger–Keldysh formalism, we examine open extensions of three well-studied theories: the superfluid, Maxwell theory, and Einstein gravity. In gauge and gravitational theories, open terms that break advanced symmetries while preserving physical ones are not automatically consistent; they are allowed only if they lead to deformed identities among the equations of motion. We explicitly construct such a term in open gravity and show that it leads to a consistent deformation of the diffeomorphism identities.

}

\begin{document}

\maketitle
\flushbottom

\section{Introduction}

Understanding open quantum and classical systems is a central challenge in high-energy theory, condensed matter, and cosmology. When a system interacts with an environment, it typically exhibits non-unitary dynamics, which manifests itself as dissipation and stochastic noise, due to the exchange of energy and information. As standard quantum field theory techniques cannot capture these effects, one has to work within the Schwinger-Keldysh (SK) formalism \cite{Schwinger:1960qe,Keldysh:1964ud,Feynman:1963fq,Sieberer:2015hba,Crossley:2015evo,Glorioso:2017fpd,Liu:2018kfw,Haehl:2016pec} (see also \cite{Galley:2012hx,Galley:2014wla} for a classical formulation of nonconservative systems). This framework, which includes doubling the dynamical fields to describe evolution along a closed-time contour, enables the computation of in-in expectation values and the construction of effective actions for out-of-equilibrium processes. Such actions typically possess two types of symmetries: physical (diagonal) symmetries transform both fields identically in the two branches, while advanced symmetries act oppositely on the two field copies \cite{Sieberer:2015svu}.

Building on the foundational work of \cite{Crossley:2015evo,Liu:2018kfw} on dissipative effective field theories (EFTs), recent studies have applied these techniques in a variety of physical settings: time-translational symmetry breaking in flat spacetime \cite{Hongo:2018ant}, open EFTs of inflation in cosmology \cite{Salcedo:2024smn,Salcedo:2025ezu}, damping of pseudo-Goldstone modes \cite{Delacretaz:2021qqu,Armas:2021vku,Baggioli:2023tlc}, coset constructions with softly broken symmetries \cite{Akyuz:2023lsm,Akyuz:2025bco} and gravitational EFTs \cite{Lau:2024mqm}. A central theme in all previous works is the explicit breaking of the advanced symmetry while the physical symmetry remains intact. Interestingly, in a recent open extension of Maxwell theory, it was shown that the effect of breaking the advanced symmetry only deforms the constraints of the theory \cite{Salcedo:2024nex}; this is a consequence of invariance of the part of the action that mixes physical and advanced fields under a deformed gauge symmetry of the advanced field. Because of this new symmetry, the deterministic EOMs remain consistent in the sense that the number of independent equations matches the number of gauge-fixed variables. This raises a question: is this a universal behaviour of open EFTs in gauge or gravity theories? An alternative and equivalent formulation of the question is the following: are the EOMs of an open extension in such theories consistent for a broad class of allowed open terms?

In this work, we investigate this issue through representative examples of open effective field theories involving gauge fields and gravity. Starting from the open EFT of a superfluid, we show how an emergent average current conservation arises. We then consider an open extension of Maxwell theory, formulated in terms of higher-form variables and coupled to background fields (see, e.g.~\cite{Gaiotto:2014kfa,Armas:2023tyx}), and demonstrate that the deterministic equations satisfy a deformed constraint structure, recovering the deformed constraints of \cite{Salcedo:2024nex}. Finally, we explore gravitational analogues by analyzing open extensions of the Einstein–Hilbert action and show that consistency restricts the allowed form of open operators that can yield deformed diffeomorphism identities.

The organisation of this work is as follows. In sec.~\ref{sec:open_gauge} we first examine the open superfluid action and then construct an open extension of Maxwell theory within the higher-form formulation. In sec.~\ref{sec:gravity} we explore open extensions of the Einstein–Hilbert action and examine when identities exist between the equations of motion to guarantee consistency of the problem. Finally, in sec.~\ref{sec:conclusions} we summarize our main findings.


\section{Open Maxwell theory} \label{sec:open_gauge}

\subsection{The open superfluid action revisited}  \label{sec:superfluid}

To understand why an open extension of Maxwell theory works, we first consider the superfluid case. Its action at non-zero temperature is
\be 
\begin{aligned}
S_{\rm cl} = \int \left(   B_{\rm a 0} B_{ 0} - c_{\rm s}^{2} B_{{\rm a} i} B_i  
+\dots \right) 
\end{aligned} \, ,
\ee
where $B_{\mu} \equiv \partial_\mu \phi + A_{\mu}$, $A_{\mu}$ is a background field and dots denote second- and higher-order corrections that describe dissipation and noise. Here, as usual, we assume a unit covector $u_{\mu} = (-1,0,0,0)$ parameterizing the direction of dissipation. The presence of background fields makes the action manifestly invariant under gauge transformations $\phi \rightarrow \phi + \lambda(x)$ for both physical and advanced fields, as well as $A_{\mu} \rightarrow A_{\mu} - \partial_{\mu}\lambda(x)$. For this system, we can define the ``off-shell hydrodynamic'' current from \cite{Liu:2018kfw}
\be
J^{\mu}_{\rm off} \equiv {\delta S \over \delta A_{\rm a \mu} } = B^{\mu} + \dots \, .
\ee
Invariance of the generating functional under the two background gauge transformations yields the Ward identity for current conservation
\be
\langle \partial_\mu J^{\mu} \rangle_{A_{\rm a}=0} = 0 \, ,
\ee
where all advanced sources have been set to zero.

To describe a situation where charge leaks into an environment we typically add terms that break current conservation. At the level of the SK action this can be done by introducing terms that break the advanced symmetry such as $- \Gamma \phi_{\rm a} B_{0}$ (see e.g.~\cite{Hongo:2018ant,Delacretaz:2021qqu,Akyuz:2025bco}) and a generic noise $i \beta \phi_{\rm a}^2$, whose coefficient $\beta$ becomes proportional to the relaxation coefficient $\Gamma$ if the KMS condition \cite{Martin:1959jp,Kubo:1966fyg} is imposed.\footnote{For the derivative expansion to make sense $\Gamma$ should scale at least as $ \mathcal{O}(\partial)$.} 
Keeping the same definition for the off-shell current, we find that to linear order the EOM becomes deformed and current conservation is broken only through noise effects
\be
\partial_{\mu} B^{\mu} - \Gamma u_{\mu} B^{\mu} + i \beta \phi_{\rm a} + \mathcal{O} \left(\partial^2 \right) = 0 \, .
\ee 
The previous equation can also be written as 
\be \label{eq:total_current}
e^{-\Gamma t}\partial_{\mu} \left( e^{\Gamma t} B^{\mu}  \right) + i \beta \phi_{\rm a} +  \mathcal{O} \left(\partial^2 \right)  = 0 \, ,
\ee
highlighting the current suppression due to dissipation. The Dyson-Schwinger equation with $\langle \phi_{\rm a} \rangle = 0$ yields the expectation value of the ``total current'' conservation
\be \label{eq:average_current}
\langle \partial_{\mu} (e^{\Gamma t} B^{\mu}  )\rangle_{A_{\rm a}=0} = 0 \, ,
\ee
therefore, it is no longer conserved at the operator level, but only on average, which is a typical property of open systems \cite{Sieberer:2015svu}. 

The previous equation can also be understood as a consequence of invariance of the deterministic part of the action under advanced deformed transformations. Setting the background field to zero we observe that the deterministic part, written as
\be
S_{\rm det} = \int  \left(\partial_{\mu} \phia + \Gamma u_{\mu} \phia \right) \partial^{\mu } \phi + \mathcal{O}(\phia \partial^3\phi )\, ,
\ee
remains invariant for time-dependent shifts 
\be \label{eq:global_lambda}
\phia \rightarrow \phia + \Lambda(t) \, ,
\ee
by a function $\Lambda(t)$ satisfying
\be \label{eq:lambda}
\dot{\Lambda} - \Gamma \Lambda 
= 0 \, .
\ee
Note that for $\Gamma=0$ the solution for $\Lambda$ is a constant, as expected. This invariance can be extended to an arbitrary derivative order by an appropriate redefinition of the various higher-order coefficients. Interestingly, gauging this time-dependent shift 
by adding a background field $A$ implies the following transformation for the background field
\be
\delta A_{\rm a \mu} = - (\partial_{\mu} \lambda + \Gamma u_{\mu} \lambda ) \, ,
\ee
for an arbitrary function $\lambda(x)$, which is the emergent symmetry of the open EFT for light found in \cite{Salcedo:2024nex}; here we re-derived it as the gauge version of the time-dependent shift that leaves the deterministic part of the superfluid action invariant to leading order.

The open coupling $-\Gamma \phia B_0$ can also be interpreted as the momentum coupling $-\Gamma \phia \pi$, since for the closed action the canonical momentum is $\pi= {\partial L \over \partial \dot{\phi}} = B_0$; this observation will be important later when we discuss gauge fields. Having examined the superfluid case, we now return to the gauge field. From the viewpoint of generalized global symmetries, the scalar field carries a conventional zero-form $U(1)_0$ symmetry while the gauge field realizes its higher-form analogue, a $U(1)_1$ symmetry acting via shifts by a closed one-form, $\phi \rightarrow \phi + \omega$ with $\ud \omega =0$. In the following subsection, we will systematically construct the action of open electromagnetism using the formulation of \cite{Armas:2023tyx}. We will couple the one-form gauge field to a two-form background field and then break the advanced symmetry to write open terms. Setting the background fields to zero and using the covariant spacetime formulation reproduces the action of \cite{Salcedo:2024nex} in the leading derivative expansion.

\subsection{The open action with background fields} \label{subsec:background_fields}

To keep the analogy with the scalar field, we denote the gauge field by $\phi$ from which we can construct its field-strength tensor $\ud \phi$. We start with the closed SK action that describes a spontaneously broken higher-form $U(1)_{1}$ symmetry at finite temperature 
\be \label{eq:closed_action}
\begin{aligned}
S_{\rm cl} = \int &~ c_1 \,   \xi_{\rm a} \wedge \star \xi + c_2 \,   \ib \xi_{\rm a} \wedge \star \ib \xi + p_1 \, \xi_{\rm a} \wedge \xi  + p_2 \, \ud( \ib \xi_{\rm a}) \wedge \xi  \\ 
&+ \phi_{\rm a} \wedge \star j + \gamma_1~  u \wedge \phia \wedge \xi   \, ,
\end{aligned}
\ee
where 
\be
\xi \equiv \ud \phi + A \, ,
\ee
$A$ is a two-form background field, $j$ is an external current, $u$ is the one-form dual to the dissipation vector $\beta$. Assuming that $\beta$ is a time-like unit vector and that $u$ is a closed one-form
\be \label{eq:properties_u}
\ib u =-1 \, , \qquad \ud u=0 \, ,
\ee
we find $\mLb u=0$; the previous properties are trivially satisfied for the common choice $u= -\ud t$. The parity-violating terms $p_1$ and $p_2$ become total divergences in the absence of background fields. The action - minus the last two terms - is manifestly invariant under the $U(1)_1^{\rm gauge}$ transformations
\be
\phi \rightarrow \phi + \Lambda \, , \quad A \rightarrow A + \ud \Lambda \, ,
\ee
of both advanced/physical fields.\footnote{Note that the terms in the second line of \eqref{eq:closed_action} are not gauge invariant even though they are part of the closed action because of their coupling to background fields. The full action can be made invariant under $U(1)_1^{\rm gauge}$ transformations by introducing a second background field to cancel the variation of naked $\phia$ terms \cite{Armas:2023tyx}.}
 
Breaking the advanced symmetry allows open terms with naked $\phi_{\rm a}$ and its projection along the thermal vector $\ib \phi_{\rm a}$. To linear order the latter necessarily appears with a derivative $ \ud \ib \phi_{\rm a}$ (because they are contracted or wedged with $\xi$), and using Cartan's identity $\mLb = \ib \ud + \ud \ib $ we can trade them for terms with the Lie derivative or $\ib \ud \phi_{\rm a}$ (which already appears in the nondissipative theory). The dissipative part of the action up to linear terms in $\phi_{\rm a}$ and with at most one derivative acting on $\phi_{\rm a}$ includes the following terms
\be 
\begin{aligned} \label{eq:s_dissipative}
S_{\rm dis} \supset \int & \left\{  u \wedge  \left[\Gamma_1 \phi_{\rm a} + \Gamma_2 \mLb \phi_{\rm a} \right]\wedge \star \xi   
+ \gamma_2 \, u \wedge  \mLb \phi_{\rm a}  \wedge \xi \right\} \, .
\end{aligned} 
\ee
We can group the terms involving $\Gamma_{1}$ and $\Gamma_{2}$ by introducing the deformed operator
\be
\mathcal{D} \equiv \ud + \Gamma_1 u\wedge + \Gamma_2 u\wedge \mLb \, ,
\ee
from which a modified field-strength tensor follows 
\be
\mathcal{D} \phi_{\rm a}  = \ud \phi_{\rm a} + \Gamma_1 u \wedge \phi_{\rm a} + \Gamma_2 u \wedge \mLb  \phi_{\rm a} \, ,
\ee
and the combination
\be
f_{\rm a} \equiv \mathcal{D} \phi_{\rm a}  + A_{\rm a}  \, .
\ee
Using the assumed properties of $u$ (see eq.~\eqref{eq:properties_u}) and $[\mLb,\ud]=0$, the deformed operator satisfies by construction $\mathcal{D}^2 = 0$. 
Using the previous definitions, redefining some  coefficients and performing integrations by parts, the action is neatly written as
\be \label{eq:open_action}
S_{\rm open} = S_{\rm dyn}[\mD \phi_{\rm a},\ud \phi,A]  + S_{\rm b}[\phi_{\rm a},A] + S_{\rm n}[\phi_{\rm a}] \, ,
\ee
where we defined the dynamical part (the one depending on the physical gauge field $\phi$)
\be
S_{\rm dyn}[\mD \phi_{\rm a},\ud \phi,A] \equiv \int \left( -c_1 \,  f_{\rm a} \wedge \star \xi + c_2 \,   \ib(f_{\rm a}) \wedge \star \ib \xi  + \gamma_1 \, f_{\rm a} \wedge \xi + \gamma_2 \, u \wedge \ib f_{\rm a} \wedge \xi \right) \, ,
\ee
and the non-dynamical part of the action, which further splits to the part depending on background quantities, explicitly written as
\be
S_{\rm b}[\phi_{\rm a},A] \equiv   \, p_1 \, \phi_{\rm a} \wedge \ud A  + p_2 \, \ib \ud \phi_{\rm a} \wedge \ud A +  p_3 \, u \wedge  \ib\phi_{\rm a} \wedge \ud  A  + \phi_{\rm a} \wedge \star j \, ,
\ee
and the noise part which we consider unconstrained
\be
S_{\rm n}[\phi_{\rm a}] \equiv  \int \phi_{\rm a} \wedge \mathcal{N} \phi_{\rm a} \, ,
\ee
where $\mathcal{N} $ is a differential operator.
Every term involving the background field $A$ is invariant under the (physical) background gauge transformations $A \rightarrow A + \ud \Lambda$. 

Performing integrations by parts on the dynamical part, we can schematically write it as
\be
S_{\rm dyn} = \int \phi_{\rm a} \wedge \mathcal{E} \, ,
\ee
with $\mathcal{E}$ denoting the Euler-Lagrange expressions. The latter contains the adjoint operator $\mD^\dag$ which is  
\be
\mD^{\dag} = \ud - \Gamma_1 u \wedge + \Gamma_2 u\wedge \mLb \, .
\ee
More specifically, we find
\be
\mathcal{E} = \mD^{\dag} \star \left\{ - c_1 \,  \xi + c_2 \,   u \wedge \ib \xi  - \gamma_1 \star \xi - \gamma_2 \star \left[u \wedge \ib (\star \xi) \right] \right\} \, .
\ee
To examine the constraint structure of the theory we focus on the EOM of the physical field: $\mathcal{E}=0$. Because $\mD^\dag$ is a nilpotent operator, we immediately obtain the deformed identity
\be \label{eq:deformed_maxwell}
\mathcal{D}^\dagger \mathcal{E} = 0 \, ,
\ee
that generalizes $\ud \mathcal{E}_{\rm cl} = 0$ of the closed theory.\footnote{In the closed theory this identity is trivially satisfied because the EOM is precisely the current conservation equation $\mathcal{E}_{\rm cl} = \ud \star \mathcal{J} $.} This identity guarantees that only three of the four equations are independent, in agreement with the number of gauge-invariant combinations (see also App.~\ref{app:gauge_invariant}). Upon solving the resulting constraint equation, one finds that the theory propagates only two degrees of freedom, the transverse polarizations of the photon, and therefore that this open-EFT extension is dynamically consistent. Similarly to the superfluid action, the leading open term is a coupling to the canonical momentum of closed theory, which is the electric field $\pi = \ib \xi$. This term generates the dissipative coupling in \eqref{eq:s_dissipative} and at the level of effective equations of motion it implements Ohmic dissipation $J_{\rm dis} = \sigma E$ (see e.g.~\cite{Tong:2016kpv}).

The previous identity can also be obtained as the result of invariance of the dynamical part of the action under deformed gauge transformations of the advanced field. Notice that the dependence on $\phi_{\rm a}$ in $S_{\rm dyn}$ is exclusively through the modified field-strength two-form $\mD \phi_{\rm a}$, thus, transformations that leave it invariant will be equivalent to identities between the EOMs (see app.~\ref{subsec:def_gauge} for an interpretation of this deformed gauge symmetry in the closed theory). In parallel with Maxwell theory, these transformations involve shifts by $\mD$-exact one-forms, i.e.~those satisfying $\omega_{\rm a} = \mD \Lambda$ for some scalar function $\Lambda(x)$ (because then $\delta \mD \phia = \mD^2 \Lambda=0$). The global subgroup for which the gauge field does not change, i.e.~the non-trivial $\Lambda$ satisfying 
$\mathcal{D}\Lambda=0$, is exactly the deformed time-dependent shift \eqref{eq:global_lambda} found in sec.~\ref{sec:superfluid} (up to a possible rescaling). Having found identities between the EOMs, we briefly discuss what these imply for the noise part.

Including noise, the equation of motion with respect to $\phi_{\rm a}$ becomes 
\be
\mathcal{E}  + {\delta S_{\rm b} \over \delta \phi_{\rm a} } + \Xi = 0 \, .
\ee
where $\Xi \equiv \left(\mathcal{N} + \mathcal{N}^\dagger \right) \phi_{\rm a}$. 
Acting with $\mathcal{D}^\dagger$ on the EOM gives a constraint on the advanced components 
\be
\mathcal{D}^\dagger  \left( {\delta S_{\rm b} \over \delta \phi_{\rm a} }  + \Xi \right) = 0 \, ,
\ee
which generalizes the noise constraint of \cite{Salcedo:2024nex} in the presence of background fields. This equation reduces the number of advanced components that couple to the dynamical fields from four to three to match the three components of the gauge-fixed physical field in the action.

\section{Deformed identities in open gravity} \label{sec:gravity}

The case of dissipative gravity is qualitatively different from that of gauge fields for a variety of reasons. First, the standard formulation in the literature is valid in the semi-classical limit by treating the advanced field as a fluctuation field \cite{Crossley:2015evo,Liu:2018kfw}, which implies the following form of advanced diffeomorphisms: $\delta g_{\rm a}^{\mu \nu} = \mathcal{L}_{\xi} g^{\mu \nu}$. This can be seen by writing the simplest closed action constructed from the Einstein-Hilbert term and noticing its similarities to the variation of the action
\be
S_{\rm cl} = \int \sqrt{-g} \ud^4 x G_{\mu \nu}  g_{\rm a}^{\mu \nu}  \sim  \int \sqrt{-g} \ud^4 x G_{\mu \nu}  \delta g^{\mu \nu} \,.
\ee
The latter action has the usual diffeomorphism symmetry $\delta g^{\mu\nu}=\mathcal{L}_{\xi} g^{\mu\nu}$ and the advanced symmetry mentioned above follows. The EOM of the closed theory satisfies the Bianchi identity $\nabla^{\mu} G_{\mu \nu}=0$ which ensures that the number of gauge-fixed variables matches the number of independent equations. Note also that physical diffeomorphisms affect the advanced metric, which is now viewed as a tensor living on the physical spacetime. Second, when introducing a dissipation vector, we reduce the symmetries of diffeomorphisms to a subgroup involving reparameterizations of the hyperspace orthogonal to that vector. This may generate new degrees of freedom or alter the constraint structure of the theory.

To capture the effects of an unknown environment we start with a modification of Einstein's equations
\be \label{eq:modified_equation}
E_{\mu \nu} = G_{\mu \nu} + \Delta_{\mu \nu} \, ,
\ee
and encode diffeomorphism breaking via a non-covariantly conserved $\Delta_{\mu \nu}$ leading to a non-zero source $J_{\nu}$, 
\be \label{eq:divergence_equation}
\nabla_{\mu} E^{\mu}_{~\nu} = \nabla_{\mu} \Delta^\mu_{~\nu} \equiv J_{\nu} \, .
\ee 
We also consider a unit vector parameterizing open effects, which naturally provides a foliation of spacetime, and thus we identify it with the normal vector of the ADM decomposition. Treating the covector as fixed introduces a preferred direction, and hence we expect invariance of the theory only for diffeomorphisms of the induced hypersurface, similar to the EFT of inflation \cite{Cheung:2007st}. Using the previous recipe,  the deformed identity becomes 
\be \label{eq:deformed_identity}
P^\nu_i \nabla^\mu E_{\mu \nu} = \alpha P^\nu_i n^\mu E_{\mu \nu} + \dots \, .
\ee
where $P^{\mu \nu} \equiv \delta^\mu_\nu + n^\mu n_\nu$ is the projector and we expect the identity to be satisfied only for diffeomorphisms tangential to the hypersurface orthogonal to $n_{\mu}$. We now turn our attention to $\Delta_{\mu \nu}$. 

One might expect that every foliation-preserving term would constitute a viable choice, similar to Maxwell theory in 4D; there, we saw that with the assumed properties of the unit vector, all allowed open terms were consistent with the dynamics. Here, however, we are limited by the number of identities in the ADM formalism and more generally in gravity, and therefore, we expect a small number of non-trivial open terms.\footnote{There are certain types of open terms which are trivial in the sense that they do not alter the propagating equations (see app.~\ref{app:trivial}).} One such term involves the extrinsic curvature ($K_{\mu \nu} \equiv P^\alpha_\mu  \nabla_\alpha n_\beta$) through the Codazzi equation, frequently encountered in the Hamiltonian formulation of general relativity
\be
G_{\mu \nu}n^\mu P^\nu_i = D^j(K_{ij} - K P_{ij}) \, ,
\ee
where $D_i$ is the covariant derivative with respect to the induced metric ($D_i \equiv P^\mu_i \nabla_\mu$). The previous relation suggests an open term constructed from the combination $\tilde{K}_{\mu \nu} \equiv K_{\mu \nu} - K P_{\mu \nu}$. However, this combination is not yet complete; taking the divergence of the previous term and projecting on the hypersurface yields
\be \label{eq:projected_k}
P^\nu_i \nabla^\mu \tilde{K}_{\mu \nu}  = D^j \tilde{K}_{ij} + a^j \tilde{K}_{ij} \, ,
\ee
where $a_\mu \equiv n^{\nu} \nabla_{\nu} n_{\mu}$ is the covariant acceleration defined from $\nabla_\mu n_\nu = K_{\mu \nu} - n_\mu a_\nu$. To cancel the second term in \eqref{eq:projected_k}, we note that in the ADM variables, the covariant acceleration is given as the covariant derivative with respect to the induced metric $a_i = D_i(\log \mathcal{N})$ of the lapse function. Expressing the acceleration in terms of the lapse function, dividing eq.~\eqref{eq:projected_k} by the lapse and using the chain rule, we obtain the exact identity 
\be \label{eq:divergence_extr_curv}
P^\nu_i \nabla^\mu \left( {\tilde{K}_{\mu \nu} \over \mathcal{N}} \right) = {1 \over \mathcal{N}} D^j \tilde{K}_{ij} =  {1 \over \mathcal{N}} P^\nu_i G_{\mu \nu} n^\mu\, .
\ee
Finally, if our EOM satisfies $E_{\mu \nu}n^\mu P^\nu_i  = G_{\mu \nu} n^\mu P^\nu_i $ we obtain a first \textit{non-trivial  identity} between EOMs in open gravity
\be \label{eq:divergence_identity}
P^\nu_i \nabla^\mu E_{\mu \nu} = {\gamma \over \mathcal{N}} E_{\mu \nu}n^\mu P^\nu_i\, .
\ee
When noise is included, the previous identity also implies a set of three noise constraints. An example of a theory that satisfies the identity \eqref{eq:divergence_identity} is the open EFT of inflation (see the subsequent publication by the author of this work \cite{per}). Using the relation between the lapse and the normal vector
\be
n_\mu = -\mathcal{N} \partial_\mu \theta \,, 
\ee
for a scalar function $\theta$, we can write the deformed identity as
\be
P^\nu_i  \left(\nabla^\mu E_{\mu \nu} + \gamma  E_{\mu \nu} \partial^\mu \theta \right) = e^{-\gamma \theta}P^\nu_i \nabla^\mu \left( e^{\gamma \theta} E_{\mu \nu} \right)  = 0 \, ,
\ee
in analogy to \eqref{eq:total_current}, which shares similarities to a gravity theory with a dilaton coupling. 

In terms of canonical variables, the previous open term is related to the canonical momentum in the ADM decomposition
\be \label{eq:momentum_adm}
{1\over \mathcal{N}}(K_{ij} - K P_{ij}) = {\pi_{ij} \over \mathcal{N} \sqrt{\gamma} }  \, , 
\ee
and therefore generates the following term in the SK action
\be 
\int \ud^4 x \, \pi_{ij} \gamma_{\rm a }^{ij} \, ,
\ee
where $\gamma_{ij}$ is the induced metric, in complete analogy to the superfluid and Maxwell theory.\footnote{Notice that the previous requires identifying $\pi_{ij}$ with the canonical momentum of general relativity and not just ${\partial L \over \partial \dot{h}^{ij}}$ of the full closed theory in the unitary gauge, which would add contributions from terms such as $\delta K \delta g^{00}$.}

\section{Summary} \label{sec:conclusions}

How is dissipation consistently introduced in a system with gauge or diffeomorphism symmetries? When physical symmetries are preserved, operators that break the advanced symmetry can be consistently included provided that they respect a (possibly deformed) constraint structure.  We observed that a broad and physically motivated class of consistent dissipative terms can be written as momentum couplings of the form $ \phia \wedge \star \pi$ (or $A_{\rm a}^i \pi_i $ in tensor notation) in Maxwell theory and $\pi_{ij} \gamma_{\rm a}^{ij}$ in gravity,  with $\pi$ denoting the canonical momentum of the free theory. These terms deform the off-shell identities of the theory, guaranteeing the correct number of propagating degrees of freedom without overconstraining the system.

This work lays the foundations for a systematic investigation of dissipative effects in gravity. An example of a theory satisfying a deformed diffeomorphism identity found in sec.~\ref{sec:gravity} is the open EFT of inflation (studied in a subsequent publication by the author of this work \cite{per}) which extends previous constructions formulated in the decoupling limit \cite{LopezNacir:2011kk,Creminelli:2023aly,Salcedo:2024nex}. A promising direction for future work is to develop open extensions of effective field theories of dark energy \cite{Gubitosi:2012hu,Gleyzes:2013ooa} and to explore whether dissipative couplings give rise to distinctive observational signatures.

\acknowledgments
I thank Vaios Ziogas for initial collaboration and many insightful discussions on higher forms and open hydrodynamics, and Thomas Colas for useful correspondence. I also thank the Korea Institute for Advanced Study for its hospitality. I am supported in part by the National research foundation of Korea Grant 2-2024-1058-001-2.

\begin{appendix}

\section{Some remarks on open Maxwell theory}
\subsection{Gauge-invariant variables} \label{app:gauge_invariant}
To gain some insight on the constraint structure of the open theory, we examine how degrees of freedom of a gauge field become manifest in the longitudinal/transverse decomposition. Recall the free Maxwell action and its EOM 
\begin{align}
S &= -{1\over 4} \int F^2 \, , \\
\mathcal{E}_{\nu} &\equiv \partial^{\mu}F_{\mu \nu} =  0 \, .
\end{align}
Decomposing the spatial part of the gauge field (assuming appropriate boundary conditions) as
\be
A_{i} = \partial_{i} \phi + A^{\rm tr}_{i} \, ,
\ee
the time and space components of the Euler-Lagrange expressions are written as
\begin{align}
\mathcal{E}_0 &= \partial^{j} \partial_{j} (A_0  - \dot{\phi}) \, , \\
\mathcal{E}_i &=  \partial_{i} \partial_t(A_0 - \dot{\phi})  + \Box A_{i}^{\rm tr}   \, .
\end{align}
Notice that the four components of the gauge field appear in two gauge-invariant combinations, $A_0 - \dot{\phi}$ and $A_{i}^{\rm tr}$, which represent 3 independent variables. Moreover, the two equations combine to 
\be
\partial^\mu \mathcal{E}_\mu = 0 \, ,
\ee
which is trivially valid due to the antisymmetry of $F_{\mu \nu}$.

For the leading open terms, the EOM of the free part is modified to
\be
\tilde{\mathcal{E}}_{\nu} = \left( \partial^{\mu} - \Gamma u^{\mu}\right) F_{\mu \nu} = 0 \, ,
\ee
which alters the spatial component of the EOM to
\be
(\partial_t + \Gamma )\partial_{i} (A_0 - \dot{\phi})  + \Box A_{i}^{\rm tr}  - \Gamma  \dot{A}_i^{\tr} = 0 \, .
\ee
Although the EOM and thus the system's dynamics has changed, the gauge-invariant combinations are the same; any gauge-fixing condition that works in the closed theory (e.g.~Coulomb gauge) works here as well. The EOM satisfies the deformed identity 
\be \label{eq:deformed_em}
\left(\partial^{\mu} - \Gamma u^{\mu}\right) \mathcal{\tilde{E}}_{\mu} = 0 \, ,
\ee
which is eq.~\eqref{eq:deformed_maxwell} in tensor notation. This identity is trivially valid due to the antisymmetry of $F_{\mu \nu}$, and its existence guarantees a consistent EOM.

\subsection{Deformed gauge symmetry} \label{subsec:def_gauge}
In this section we will show that the deformed gauge transformation of the dynamical part of the action studied in sec.~\ref{subsec:background_fields} can be a legitimate symmetry of a non-dissipative Maxwell-like theory without Lorentz invariance. This is in sharp contrast with gravity where deformed diffeomorphisms are incompatible with a closed action. To obtain such a theory we assume a current $J^{\mu}$ coupled to a gauge field $A_{\mu}$ and assume that the current satisfies a deformed conservation law of the form
\be \label{eq:current_non_conservation}
 l_{\mu} J^{\mu} = 0 \, ,
\ee
with $l_{\mu}$ some differential operator. As a first step, we impose this deformation at the level of the action and hence it should be related to the gauge invariance of a gauge field.\footnote{Although this is not a strict requirement it will facilitate the subsequent discussion. When relaxing this assumption, current deformation holds on-shell (see eq.~\eqref{eq:current_deformation}).} The coupling of this current to a gauge field, $J^{\mu}A_{\mu}$, yields this identity provided that the gauge field transforms as
\be
J^{\mu} \delta A_{\mu} = J^{\mu} \mathcal{D}_{\mu} \phi \, ,
\ee
for some differential operator $\mathcal{D}_{\mu} =  l_{\mu}^{\dag}$ that generalizes the ordinary partial derivative $\partial_{\mu}$. 

We proceed to construct an action that is invariant under the deformed gauge symmetry
\be
\delta A_{\mu}= \mathcal{D}_{\mu } \phi \, .
\ee
By requiring $\mathcal{D}_{\mu}$ to be a commuting differential operator, i.e.~$[\mathcal{D}_{\mu},\mathcal{D}_{\nu}]=0$,\footnote{This is equivalent to requiring the deformed derivative operator to satisfy $\mathcal{D}^2=0$ when it acts on differential forms in complete analogy to the exterior derivative $\ud^2=0$ (see also sec.~\ref{subsec:background_fields}).} we can define a modified field-strength tensor
\be
f_{\mu\nu} \equiv \mathcal{D}_{\mu} A_{\nu} -\mathcal{D}_{\nu} A_{\mu} \, ,
\ee
and build a Lagrangian that is invariant under this deformed symmetry
\be
\mathcal{L} = - {1\over 4} f^2 - J^{\mu} A_{\mu}\, ,
\ee
with equations of motion 
\be \label{eq:mod_Maxwell}
\mathcal{D}_{\mu}^{\dag} f^{\mu \nu} = J^{\nu} \, .
\ee
Deformed gauge invariance implies the Noether identity
\be
\mathcal{D}_{\mu}^{\dag} \mathcal{D}_{\nu}^{\dag} f^{\mu \nu} = 0 \, ,
\ee
and through the EOM the on-shell deformed current conservation
\be \label{eq:current_deformation}
\mathcal{D}_{\nu}^{\dag} J^{\nu} = 0 \, .
\ee
For the choice of the deformed operator of sec.~\ref{sec:superfluid}, $\mathcal{D}_{\mu} = \partial_{\mu} + \Gamma u_{\mu}$, for some spacetime covector $u_{\mu}$, the commutativity property of $\mathcal{D}$ requires $u$ to satisfy
\be
\partial_{\mu} u_{\nu} - \partial_{\nu} u_{\mu} = 0\, ,
\ee
or to be given as the gradient of a function, i.e.~$u_{\mu} =\partial_{\mu}\psi$. Of course, this includes the simplest choice of a constant timelike $u_{\mu}$. The equations of motion of this theory have a structure similar to Maxwell theory, i.e.~$A_0$ is still a constraint field, since there are no second-order time derivatives acting on it, and using gauge invariance one of the spatial components of $A_{\mu}$ can be removed. Thus, this theory describes two propagating degrees of freedom, albeit with no Lorentz symmetry.

\section{Trivial open terms} \label{app:trivial}

There are certain types of open terms generated by taking projections of the EOMs along or orthogonal the dissipation vector. Although these are not derived from a single-copy action, they are trivial in the sense that they do not alter the EOMs apart from a simple rescaling. This can be easily shown in the open Maxwell theory, where we can construct one such independent projection
\be
\Delta_{\mu} = \gamma u^\nu \mathcal{E}_\nu u_\mu \, ,
\ee
and it is a simple exercise to show that the temporal and spatial parts of the EOM remain unchanged. Because the extra term is constructed by projecting along the unit vector, we find for the total EOM $\tilde{\mathcal{E}}_{\mu} = \mathcal{E}_\mu + \Delta_\mu$
\be
u^\mu \tilde{\mathcal{E}}_{\nu} = (1-\gamma) u^\mu \mathcal{E}_{\nu} \, ,
\ee
which allows us to express the extra term as a projection of the total EOM
\be
\Delta_{\mu} = {\gamma \over 1-\gamma} u^\nu \tilde{\mathcal{E}}_\nu  u_{\mu} \, .
\ee
Finally, acting with a derivative on the EOM we obtain
\be
\partial^\mu \tilde{\mathcal{E}}_\mu = {\gamma \over 1-\gamma}  u^{\mu} \partial_{\mu} (u^\nu \tilde{\mathcal{E}}_\nu )  \, .
\ee
where we used the fact that $\partial^{\mu}\mathcal{E}_{\mu} = 0$ for the closed theory. Even though we constructed a non-trivial identity between the EOMs, this extension is trivial because it does not alter the EOM of propagating degrees of freedom.

An analogous situation holds in gravity. For the simplest term that is built entirely from the metric and curvature tensor,
\be \label{eq:delta_1}
\Delta_{\mu \nu} = \Gamma  R g_{\mu \nu} \, ,
\ee
we can combine the trace of \eqref{eq:modified_equation} 
\be
E = (4 \Gamma -1 ) R \, ,
\ee
with the divergence equation \eqref{eq:modified_equation} to find the deformed identity
\be
\nabla_{\mu } E^{\mu}_{~\nu}  = { \Gamma \over 4 \Gamma -1} \left(E_{\alpha \beta}g^{\alpha \beta}\right)_{,\nu} \, ,
\ee
for $\Gamma \neq 1/4$. Even though this appears as a non-trivial identity between the EOMs, the open term we added is just the trace of the EOM and, therefore, it does not alter the equations of propagating degrees of freedom, which in this case are the two polarizations of the graviton. 

Introducing additional structure in our theory allows for more projections of the EOM, which in our simple gravity model takes the form
\be \label{eq:delta}
\Delta_{\mu \nu} = A g_{\mu \nu} + B n_{\mu} n_{\nu} + \gamma_3 n^{\kappa} G_{ \kappa (\mu} n_{\nu)} \, ,
\ee
with
\be \label{eq:delta_conditions}
A = \Gamma_1 R + \Gamma_2 G_{\rm nn}  \, , \qquad  B = \gamma_1 R + \gamma_2 G_{\rm nn}  \, .
\ee
Note that the subcases with $\Delta_{\mu \nu} = R \, n_{\mu}n_{\nu}$ or $\Delta_{\mu \nu} = R \, P_{\mu\nu}$ have already been considered in \cite{Salcedo:2025ezu} and showed that linear perturbations around Minkowski spacetime produce identities at the linear level. The deformed identity holds, of course, at all orders and can be found by expessing $\Delta_{\mu \nu}$ in terms of $E_{\mu \nu}$. We will illustrate this for the more involved term $\gamma_3$ and the other two can be treated similarly. 
Setting $\Gamma_{1,2}$ and $\gamma_{1,2}$ to zero in eq.~\eqref{eq:delta} and contracting with $n^{\mu}n^{\mu}$ allows us to express $G_{\rm nn}$ in terms of $E_{\rm nn}$:
\be
E_{\rm nn} = (1- \gamma_3) G_{\rm nn} \, .
\ee
Taking one projection along $n^{\mu}$ yields the useful relation
\be
n^{\kappa} E_{\kappa (\mu} n_{\nu)} = {\gamma_3 \over 2} G_{\rm nn} n_{\mu} n_{\nu} + \left(1 - {\gamma_3 \over 2} \right) n^{\kappa} G_{\kappa (\mu} n_{\nu)} \, ,
\ee
hence, we obtain
\be \label{eq:delta_munu}
\Delta_{\mu \nu} =\left(1 - {\gamma_3 \over 2} \right)^{-1}  \gamma_3  \left(n^{\kappa} E_{\kappa (\mu} n_{\nu)} -  {\gamma_3 \over 2(1- \gamma_3)}E_{\rm nn} n_{\mu} n_{\nu} \right) \, ,
\ee
for $\gamma_3 \neq 1,2$. We stress again that all previous terms are trivial because they do not alter the EOM of propagating degrees of freedom.

\end{appendix}

\addcontentsline{toc}{section}{References}
\bibliographystyle{JHEP}
\bibliography{bibiOEFT.bib}

\providecommand{\href}[2]{#2}\begingroup\raggedright\begin{thebibliography}{10}

\bibitem{Schwinger:1960qe}
J.S.~Schwinger, \emph{{Brownian motion of a quantum oscillator}}, \href{https://doi.org/10.1063/1.1703727}{\emph{J. Math. Phys.} {\bfseries 2} (1961) 407}.

\bibitem{Keldysh:1964ud}
L.V.~Keldysh, \emph{{Diagram Technique for Nonequilibrium Processes}}, \href{https://doi.org/10.1142/9789811279461_0007}{\emph{Sov. Phys. JETP} {\bfseries 20} (1965) 1018}.

\bibitem{Feynman:1963fq}
R.P.~Feynman and F.L.~Vernon, Jr., \emph{{The Theory of a general quantum system interacting with a linear dissipative system}}, \href{https://doi.org/10.1016/0003-4916(63)90068-X}{\emph{Annals Phys.} {\bfseries 24} (1963) 118}.

\bibitem{Sieberer:2015hba}
L.M.~Sieberer, A.~Chiocchetta, A.~Gambassi, U.C.~T{\"a}uber and S.~Diehl, \emph{{Thermodynamic Equilibrium as a Symmetry of the Schwinger-Keldysh Action}}, \href{https://doi.org/10.1103/PhysRevB.92.134307}{\emph{Phys. Rev. B} {\bfseries 92} (2015) 134307} [\href{https://arxiv.org/abs/1505.00912}{{\ttfamily 1505.00912}}].

\bibitem{Crossley:2015evo}
M.~Crossley, P.~Glorioso and H.~Liu, \emph{{Effective field theory of dissipative fluids}}, \href{https://doi.org/10.1007/JHEP09(2017)095}{\emph{JHEP} {\bfseries 09} (2017) 095} [\href{https://arxiv.org/abs/1511.03646}{{\ttfamily 1511.03646}}].

\bibitem{Glorioso:2017fpd}
P.~Glorioso, M.~Crossley and H.~Liu, \emph{{Effective field theory of dissipative fluids (II): classical limit, dynamical KMS symmetry and entropy current}}, \href{https://doi.org/10.1007/JHEP09(2017)096}{\emph{JHEP} {\bfseries 09} (2017) 096} [\href{https://arxiv.org/abs/1701.07817}{{\ttfamily 1701.07817}}].

\bibitem{Liu:2018kfw}
H.~Liu and P.~Glorioso, \emph{{Lectures on non-equilibrium effective field theories and fluctuating hydrodynamics}}, \href{https://doi.org/10.22323/1.305.0008}{\emph{PoS} {\bfseries TASI2017} (2018) 008} [\href{https://arxiv.org/abs/1805.09331}{{\ttfamily 1805.09331}}].

\bibitem{Haehl:2016pec}
F.M.~Haehl, R.~Loganayagam and M.~Rangamani, \emph{{Schwinger-Keldysh formalism. Part I: BRST symmetries and superspace}}, \href{https://doi.org/10.1007/JHEP06(2017)069}{\emph{JHEP} {\bfseries 06} (2017) 069} [\href{https://arxiv.org/abs/1610.01940}{{\ttfamily 1610.01940}}].

\bibitem{Galley:2012hx}
C.R.~Galley, \emph{{Classical Mechanics of Nonconservative Systems}}, \href{https://doi.org/10.1103/PhysRevLett.110.174301}{\emph{Phys. Rev. Lett.} {\bfseries 110} (2013) 174301} [\href{https://arxiv.org/abs/1210.2745}{{\ttfamily 1210.2745}}].

\bibitem{Galley:2014wla}
C.R.~Galley, D.~Tsang and L.C.~Stein, \emph{{The principle of stationary nonconservative action for classical mechanics and field theories}},  \href{https://arxiv.org/abs/1412.3082}{{\ttfamily 1412.3082}}.

\bibitem{Sieberer:2015svu}
L.M.~Sieberer, M.~Buchhold and S.~Diehl, \emph{{Keldysh Field Theory for Driven Open Quantum Systems}}, \href{https://doi.org/10.1088/0034-4885/79/9/096001}{\emph{Rept. Prog. Phys.} {\bfseries 79} (2016) 096001} [\href{https://arxiv.org/abs/1512.00637}{{\ttfamily 1512.00637}}].

\bibitem{Hongo:2018ant}
M.~Hongo, S.~Kim, T.~Noumi and A.~Ota, \emph{{Effective field theory of time-translational symmetry breaking in nonequilibrium open system}}, \href{https://doi.org/10.1007/JHEP02(2019)131}{\emph{JHEP} {\bfseries 02} (2019) 131} [\href{https://arxiv.org/abs/1805.06240}{{\ttfamily 1805.06240}}].

\bibitem{Salcedo:2024smn}
S.A.~Salcedo, T.~Colas and E.~Pajer, \emph{{The open effective field theory of inflation}}, \href{https://doi.org/10.1007/JHEP10(2024)248}{\emph{JHEP} {\bfseries 10} (2024) 248} [\href{https://arxiv.org/abs/2404.15416}{{\ttfamily 2404.15416}}].

\bibitem{Salcedo:2025ezu}
S.A.~Salcedo, T.~Colas, L.~Dufner and E.~Pajer, \emph{{An Open System Approach to Gravity}},  \href{https://arxiv.org/abs/2507.03103}{{\ttfamily 2507.03103}}.

\bibitem{Delacretaz:2021qqu}
L.V.~Delacr\'etaz, B.~Gout\'eraux and V.~Ziogas, \emph{{Damping of Pseudo-Goldstone Fields}}, \href{https://doi.org/10.1103/PhysRevLett.128.141601}{\emph{Phys. Rev. Lett.} {\bfseries 128} (2022) 141601} [\href{https://arxiv.org/abs/2111.13459}{{\ttfamily 2111.13459}}].

\bibitem{Armas:2021vku}
J.~Armas, A.~Jain and R.~Lier, \emph{{Approximate symmetries, pseudo-Goldstones, and the second law of thermodynamics}}, \href{https://doi.org/10.1103/PhysRevD.108.086011}{\emph{Phys. Rev. D} {\bfseries 108} (2023) 086011} [\href{https://arxiv.org/abs/2112.14373}{{\ttfamily 2112.14373}}].

\bibitem{Baggioli:2023tlc}
M.~Baggioli, Y.~Bu and V.~Ziogas, \emph{{U(1) quasi-hydrodynamics: Schwinger-Keldysh effective field theory and holography}}, \href{https://doi.org/10.1007/JHEP09(2023)019}{\emph{JHEP} {\bfseries 09} (2023) 019} [\href{https://arxiv.org/abs/2304.14173}{{\ttfamily 2304.14173}}].

\bibitem{Akyuz:2023lsm}
C.O.~Akyuz, G.~Goon and R.~Penco, \emph{{The Schwinger-Keldysh coset construction}}, \href{https://doi.org/10.1007/JHEP06(2024)004}{\emph{JHEP} {\bfseries 06} (2024) 004} [\href{https://arxiv.org/abs/2306.17232}{{\ttfamily 2306.17232}}].

\bibitem{Akyuz:2025bco}
C.O.~Akyuz and R.~Penco, \emph{{Effective description of ajar systems with a U(1) symmetry}}, \href{https://doi.org/10.1103/mrzp-zh4v}{\emph{Phys. Rev. D} {\bfseries 112} (2025) L011901} [\href{https://arxiv.org/abs/2503.22840}{{\ttfamily 2503.22840}}].

\bibitem{Lau:2024mqm}
P.H.C.~Lau, K.~Nishii and T.~Noumi, \emph{{Gravitational EFT for dissipative open systems}}, \href{https://doi.org/10.1007/JHEP02(2025)155}{\emph{JHEP} {\bfseries 02} (2025) 155} [\href{https://arxiv.org/abs/2412.21136}{{\ttfamily 2412.21136}}].

\bibitem{Salcedo:2024nex}
S.A.~Salcedo, T.~Colas and E.~Pajer, \emph{{An Open Effective Field Theory for light in a medium}}, \href{https://doi.org/10.1007/JHEP03(2025)138}{\emph{JHEP} {\bfseries 03} (2025) 138} [\href{https://arxiv.org/abs/2412.12299}{{\ttfamily 2412.12299}}].

\bibitem{Gaiotto:2014kfa}
D.~Gaiotto, A.~Kapustin, N.~Seiberg and B.~Willett, \emph{{Generalized Global Symmetries}}, \href{https://doi.org/10.1007/JHEP02(2015)172}{\emph{JHEP} {\bfseries 02} (2015) 172} [\href{https://arxiv.org/abs/1412.5148}{{\ttfamily 1412.5148}}].

\bibitem{Armas:2023tyx}
J.~Armas and A.~Jain, \emph{{Approximate higher-form symmetries, topological defects, and dynamical phase transitions}}, \href{https://doi.org/10.1103/PhysRevD.109.045019}{\emph{Phys. Rev. D} {\bfseries 109} (2024) 045019} [\href{https://arxiv.org/abs/2301.09628}{{\ttfamily 2301.09628}}].

\bibitem{Martin:1959jp}
P.C.~Martin and J.S.~Schwinger, \emph{{Theory of many particle systems. 1.}}, \href{https://doi.org/10.1103/PhysRev.115.1342}{\emph{Phys. Rev.} {\bfseries 115} (1959) 1342}.

\bibitem{Kubo:1966fyg}
R.~Kubo, \emph{{The fluctuation-dissipation theorem}}, \href{https://doi.org/10.1088/0034-4885/29/1/306}{\emph{Rept. Prog. Phys.} {\bfseries 29} (1966) 255}.

\bibitem{Tong:2016kpv}
D.~Tong, \emph{{Lectures on the Quantum Hall Effect}},  6, 2016 [\href{https://arxiv.org/abs/1606.06687}{{\ttfamily 1606.06687}}].

\bibitem{Cheung:2007st}
C.~Cheung, P.~Creminelli, A.L.~Fitzpatrick, J.~Kaplan and L.~Senatore, \emph{{The Effective Field Theory of Inflation}}, \href{https://doi.org/10.1088/1126-6708/2008/03/014}{\emph{JHEP} {\bfseries 03} (2008) 014} [\href{https://arxiv.org/abs/0709.0293}{{\ttfamily 0709.0293}}].

\bibitem{per}
P.~Christodoulidis and J.-O.~Gong, ``Gravitational open effective field theory of inflation.'' , in preparation.

\bibitem{LopezNacir:2011kk}
D.~Lopez~Nacir, R.A.~Porto, L.~Senatore and M.~Zaldarriaga, \emph{{Dissipative effects in the Effective Field Theory of Inflation}}, \href{https://doi.org/10.1007/JHEP01(2012)075}{\emph{JHEP} {\bfseries 01} (2012) 075} [\href{https://arxiv.org/abs/1109.4192}{{\ttfamily 1109.4192}}].

\bibitem{Creminelli:2023aly}
P.~Creminelli, S.~Kumar, B.~Salehian and L.~Santoni, \emph{{Dissipative inflation via scalar production}}, \href{https://doi.org/10.1088/1475-7516/2023/08/076}{\emph{JCAP} {\bfseries 08} (2023) 076} [\href{https://arxiv.org/abs/2305.07695}{{\ttfamily 2305.07695}}].

\bibitem{Gubitosi:2012hu}
G.~Gubitosi, F.~Piazza and F.~Vernizzi, \emph{{The Effective Field Theory of Dark Energy}}, \href{https://doi.org/10.1088/1475-7516/2013/02/032}{\emph{JCAP} {\bfseries 02} (2013) 032} [\href{https://arxiv.org/abs/1210.0201}{{\ttfamily 1210.0201}}].

\bibitem{Gleyzes:2013ooa}
J.~Gleyzes, D.~Langlois, F.~Piazza and F.~Vernizzi, \emph{{Essential Building Blocks of Dark Energy}}, \href{https://doi.org/10.1088/1475-7516/2013/08/025}{\emph{JCAP} {\bfseries 08} (2013) 025} [\href{https://arxiv.org/abs/1304.4840}{{\ttfamily 1304.4840}}].

\end{thebibliography}\endgroup

\end{document}